\documentclass{aa}
\usepackage{graphicx}
\def\gtrsim{\mathrel{\hbox{\rlap{\hbox{\lower4pt\hbox{$\sim$}}}\hbox{$>$}}}}
\def\ltsim{\mathrel{\hbox{\rlap{\hbox{\lower4pt\hbox{$\sim$}}}\hbox{$<$}}}}

\def\kms{\hbox{{\rm km}\,{\rm s}$^{-1}$}}

\begin{document}
\title{A search for magnetic fields in the\\variable HgMn star $\alpha$ Andromedae\thanks{Based on data obtained using the MuSiCoS spectropolarimeter at Pic du Midi Observatory, France, from the European Southern Observatory telescopes obtained from the ESO/ST-ECF Science Archive Facility, and from the Canada-France-Hawaii Telescope (CFHT) which is operated by the National Research Council of Canada, the Institut National des Sciences de l'Univers of the Centre National de la Recherche Scientifique of France, and the University of Hawaii}}


   \author{G.A. Wade\inst{1}, M. Auri\`ere\inst{2}, S. Bagnulo\inst{5}, J.-F. Donati\inst{2}, N. Johnson\inst{1}, J.D. Landstreet\inst{4},\\F. Ligni\`eres\inst{2}, S. Marsden\inst{2}, D. Monin\inst{3}, D. Mouillet\inst{2}, F. Paletou\inst{2}, P. Petit\inst{2}, N. Toqu\'e\inst{2},\\E. Alecian\inst{6}, C. Folsom\inst{1,7}}

   \offprints{G.A. Wade, {\tt Gregg.Wade@rmc.ca}}

   \institute{Physics Dept., Royal Military College of Canada, 
   PO Box 17000, Station 'Forces', Kingston, Canada K7K 4B4
\and
   Observatoire Midi-Pyr\'en\'ees, 14 Avenue Edouard Belin, Toulouse, France
\and
   Herzberg Institute of Astrophysics, 5071 West Saanich Road, Victoria, Canada V9E 2E7 
\and
  Department of Physics \& Astronomy, The University of Western Ontario, London, Ontario, Canada, N6A 3K7
\and
   European Southern Observatory, Casilla 19001, Santiago 19, Chile
\and
Obs. de Paris LESIA, 5 place Jules Janssen, 92195 Meudon Cedex, France
\and
Department of Physics, Queen's University, Kingston, Canada K7L 3N6}
          
   \date{Received ??; accepted ??}

   \abstract{The chemically peculiar HgMn stars are a class of Bp stars which have historically been found to be both non-magnetic and non-variable. Remarkably, it has recently been demonstrated that the bright, well-studied HgMn star $\alpha$~And exhibits clear Hg~{\sc ii} line profile variations indicative of a non-uniform surface distribution of this element.}{With this work, we have conducted an extensive search for magnetic fields in the photosphere of $\alpha$~And.}{We have acquired new circular polarisation spectra with the MuSiCoS and ESPaDOnS spectropolarimeters. We have also obtained FORS1 circular polarisation spectra from the ESO Archive, and considered all previously published magnetic data. This extensive dataset has been used to systematically test for the presence of magnetic fields in the photosphere of $\alpha$~And. We have also examined the high-resolution spectra for line profile variability.}
{The polarimetric and magnetic data provide no convincing evidence for photospheric magnetic fields. The highest-S/N phase- and velocity-resolved Stokes $V$ profiles, obtained with ESPaDOnS, allow us to place a $3\sigma$ upper limit of about $100$~G on the possible
presence of any undetected pure dipolar, quadrupolar or octupolar surface magnetic fields (and just 50~G for fields with significant obliquity). We also consider and dismiss the possible existence of more complex fossil and dynamo-generated fields, and discuss the implications of these results for explaining the non-uniform surface distribution of Hg. The very high-quality ESPaDOnS spectra have allowed us to confidently detect variability of Hg~{\sc ii} $\lambda 6149$, $\lambda 5425$ and $\lambda 5677$. The profile variability of the Hg~{\sc ii} lines is strong, and similar to that of the Hg~{\sc ii}\ $\lambda 3984$ line.  On the other hand, variability of other lines (e.g. Mn, Fe) is much weaker, and appears to be attributable to orbital modulation, continuum normalisation differences and weak, variable fringing. }
{} 

   \maketitle

\section{Introduction}
 
Although Ap HgMn stars have generally been found to be both non-variable (e.g. Adelman 1998) and non-magnetic
(Shorlin et al. 2002), Adelman et al. (2002) reported convincing
evidence that the bright, well-established HgMn star $\alpha$
Andromedae A (HD 358) exhibits clear ($>30$\%) variability of the
equivalent width of the Hg~{\sc ii} $\lambda$3984 line profile.

The Hg line variations observed by Adelman et al. (2002) were
consistent with a period of $2.38236\pm 0.00011$ days, a value in good
agreement with the rigid rotation period $P=2.53\pm 0.4$~days
calculated using the measured $v\sin i$ of
$52\pm 2$ km/s, the inferred rotational axis inclination $i=74^{\rm
o}$ and the stellar radius $R=2.7\pm 0.4\ R_\odot$ (Ryabchikova et al. 1999).

Interpreting the Hg line variability as the consequence of
rotational modulation of a nonuniform photospheric abundance
distribution, Adelman et al. (2002) inverted the variations using the
Doppler Imaging technique to recover a map of the Hg surface
distribution. The derived distribution is characterised by a strong ($\sim
2.5-4.5$ dex) enhancement of the Hg abundance in the {rotational}
equatorial regions, from about $-30^{\rm o}$ to $+30^{\rm o}$
rotational latitude. The equatorial Hg enhancement appears to be
structured into a series of three spots of very high
abundance, at longitudes of $90\degr$, $170\degr$ and $270\degr$, and 
connected by ``bridges'' of somewhat lower Hg abundance.

In the atmospheres of classical Ap stars, magnetic fields seem to be a
necessary ingredient in the production of non-uniform surface
abundance distributions (see, e.g. Auri\`ere et al. 2004). Could the inferred non-uniform surface
distribution of Hg imply that $\alpha$~And is a magnetic HgMn star? Such a
discovery would run counter to the conclusions of many studies (e.g. Borra \& Landstreet 1980, Shorlin et al. 2002), and could lead to important new insights into the processes responsible for the production of atmospheric chemical peculiarities.

Longitudinal magnetic field observations of $\alpha$~And have been obtained previously by Borra \& Landstreet (1980, 5 measurements), Glagolevskij et al. (1985, 6 measurements), and Chountonov (2001, 2 measurements), with best formal error bars of about 40-50 G and with no detection of a magnetic field. Although all of these authors obtained measurements of the longitudinal magnetic field using circular polarisation, none obtained the Stokes $V$ spectrum, and none obtained a complete sampling of the 2.38-day rotational cycle. 

Given the remarkable behaviour of $\alpha$~And and the limitations of previous studies of its magnetic properties, we have decided to explore further the hypothesis that $\alpha$~And is a magnetic HgMn star. In this paper we discuss available magnetic field measurements of this star, including new high-S/N circular spectropolarisation observations, well-distributed according to the rotational ephemeris of Adelman et al. (2002), along with a detailed examination of the optical line profiles and spectrum variability.

\section{Observations}

\subsection{TBL-MuSiCoS Stokes $V$ spectra}

To search for a magnetic field in $\alpha$ And, in 2000 December and 2003 June-August, 23 high S/N Stokes $V$ spectra of $\alpha$ And were obtained using
the MuSiCoS spectropolarimeter on-board the 2m Bernard Lyot telescope at Pic du Midi Observatory. During these observing runs, many spectra of other magnetic and non-magnetic stars were obtained (e.g. Shorlin et al. 2002, Petit et al. 2004a,b, Auri\`ere et al. 2004, Johnson 2004, etc.). These results support the general accuracy of the instrument for Stokes $V$ spectropolarimetry.

 The MuSiCoS instrument consists of a table-top cross-dispersed
 \'echelle spectrograph fed via a double optical fibre directly from a
 Cassegrain-mounted polarisation analysis module.   In one
 exposure, this apparatus allows the acquisition of a 
 circular polarisation (Stokes $V$) stellar
 spectrum throughout the spectral range 4500 to 6600~\AA\ with a resolving power of
 about 35,000.  The optical characteristics of the spectrograph and
 polarisation analyser, as well as the  spectropolarimeter observing
 procedures, are described in detail by Baudrand \& B\"ohm (1992) and
 by Donati et al. (1999). 

 A complete Stokes $V$ polarimetric exposure consists of a sequence of 4
 subexposures, between which the retarder is rotated by $\pm 90\degr$.  
 This has the effect of
 exchanging the beams within the whole instrument, and in particular
 switching the positions of the two orthogonally polarised spectra on
 the CCD. This observing procedure should in principle suppress all
 first-order spurious polarisation signatures down to well below the
 noise level (Wade et al. 2000a).

 The spectra were reduced using the {\sc ESpRIT} reduction package,
 described by Donati et al. (1997), in the manner described by Wade et al.
 (2000a). The journal of observations is reported in Table 1.

\begin{table}
\begin{center}
\begin{tabular}{ccccc}
\hline
UT Date & HJD           & Phase  & $\Delta t$& S/N \\
\noalign{\smallskip}
\hline
\noalign{\smallskip}

\noalign{\smallskip}
\hline
\noalign{\smallskip}
12 dec 00 &2451891.32&   0.23769 & 696 & 440\\
24 jun 03 &2452815.65&   0.22582 & 600 & 670\\ 
26 jun 03 &2452817.65&   0.06532 & 600 & 820\\ 
04 jul 03 &2452825.64&   0.41914 & 600 & 660\\ 
07 jul 03 &2452828.65&   0.68469 & 600 & 550\\
09 jul 03 &2452830.64&   0.51790 & 600 & 710\\  
09 jul 03 &2452830.65&   0.52210 & 600 & 710\\ 
09 jul 03 &2452830.66&   0.52630 & 600 & 730\\  
10 jul 03 &2452831.65&   0.94185 & 600 & 570\\  
10 jul 03 &2452831.66&   0.94605 & 600 & 680\\  
10 jul 03 &2452831.67&   0.95025 & 600 & 560\\  
11 jul 03 &2452832.65&   0.36160 & 600 & 610\\  
12 jul 03 &2452833.60&   0.76037 & 600 & 610\\  
17 jul 03 &2452838.60&   0.85913 & 600 & 620\\ 
17 jul 03 &2452838.61&   0.86332 & 600 & 670\\ 
17 jul 03 &2452838.62&   0.86752 & 600 & 680\\  
18 jul 03 &2452839.61&   0.28308 & 600 & 710\\  
18 jul 03 &2452839.62&   0.28727 & 600 & 600\\  
19 jul 03 &2452840.59&   0.69443 & 600 & 630\\ 
19 jul 03 &2452840.60&   0.69863 & 600 & 680\\  
19 jul 03 &2452840.61&   0.70283 & 600 & 650\\  
26 jul 03 &2452847.64&   0.65368 & 600 & 610\\  
26 jul 03 &2452847.65&   0.65788 & 600 & 640\\  
26 jul 03 &2452847.66&   0.66208 & 600 & 820\\
\noalign{\smallskip}\hline\noalign{\smallskip}
22 feb 05 &2453423.70&   0.45499 & 240 & 1608\\
22 feb 05 &2453423.70&   0.45751 & 240 & 1450\\
22 feb 05 &2453423.71&   0.46003 & 240 & 1440\\
22 feb 05 &2453423.72&   0.46255 & 240 & 1450\\
18 jul 05 &2453571.07&   0.31508 &  40 & 962\\
23 aug 05 &2453607.01&   0.37557 & 120 & 1834\\
23 aug 05 &2453607.01&   0.37710 & 120 & 1922\\
23 aug 05 &2453607.02&   0.37866 & 120 & 1825\\
23 aug 05 &2453607.02&   0.38046 & 120 & 1789\\
24 aug 05 &2453607.97&   0.77786 & 120 & 1789\\
24 aug 05 &2453607.97&   0.77938 & 120 & 1859\\
24 aug 05 &2453607.98&   0.78090 & 120 & 1879\\
24 aug 05 &2453607.98&   0.78243 & 120 & 1802\\
25 aug 05 &2453609.00&   0.20979 & 120 & 1843\\
25 aug 05 &2453609.00&   0.21134 & 120 & 1823\\
25 aug 05 &2453609.00&   0.21288 & 120 & 1944\\
25 aug 05 &2453609.01&   0.21440 & 120 & 1946\\

\hline\hline\noalign{\smallskip}
\end{tabular}
\caption[]{Journal of MuSiCoS (2000 and 2003) and ESPaDOnS (2005) spectropolarimetric observations of $\alpha$ And showing UT date, HJD, phase, total exposure duration and S/N. Phases are calculated according to the ephemeris described in the text.}
\label{tab:journal}
\end{center}
\end{table}

\subsection{CFHT-ESPaDOnS Stokes $V$ spectra}

Seventeen Stokes $V$ spectra of $\alpha$~And were obtained using the new ESPaDOnS (Echelle SpectroPolarimetric Device for Observations of Stars) spectropolarimeter at the Canada-France-Hawaii Telescope (CFHT). Four were obtained on 22 Feb 2005, one was obtained on 18 July 2005, and 12 were obtained during the nights of 23-25 August 2005. During each of these missions, Stokes $V$ observations of magnetic and non-magnetic standard stars were obtained, allowing us to verify the nominal functioning of the instrument. 

 The ESPaDOnS spectropolarimeter is fundamentally similar in construction to the MuSiCoS spectropolarimeter, and allows the acquisition of a 
 circular polarisation (Stokes $V$) stellar
 spectrum throughout the spectral range 3700 to 10500~\AA\ with a resolving power of
 about 65,000.  The optical characteristics of the spectrograph and
 polarisation analyser, as well as the spectropolarimeter observing
 procedures, are described by Donati et al. (in preparation)\footnote{{ For more details about this instrument, the reader is invited to visit www.ast.obs-mip.fr/projets/espadons/espadons.html.}}.

The ESPaDOnS spectra were reduced using the Libre-ESpRIT reduction tool (Donati et al., in preparation), in a manner fundamentally similar to that described by Donati et al. (1997).

 The journal of ESPaDOnS observations is also provided in Table 1.


\subsection{VLT-FORS1 Stokes $V$ spectra}

Two FORS1 measurements of $\alpha$~And were obtained from the ESO-VLT Archive. These consist of two series of 8 Stokes $V$ subexposures, obtained in high gain mode, which were reduced and analysed using the procedures described by Bagnulo et al. (2002, 2005).

FORS1 is a low-resolution, Cassegrain-mounted spectropolarimeter. Using the GRIS600B grism in circular polarisation mode with a 0.4$\arcsec$ slit, FORS1 provides Stokes $V$ spectra with a resolving power of about 1950, covering a wavelength range from the Balmer limit to about 5900~\AA.

\subsection{UVES spectra}

A search of the ESO/ST-ECF Science Archive uncovered high-resolution,
high signal to noise ratio spectra obtained with the UVES instrument
of the ESO VLT at five different epochs, from 16 to 20 November 2001,
within the context of program ID 268.D-5738. At each observing epoch, two
observations were carried out, using both offered dichroic
modes. Almost the full wavelength interval from 3030 to 10400~\AA\ was
observed except for a few gaps, the largest of which are at
$5750-5850$~\AA\ and $8550-8650$~\AA.  In the UVES blue arm, slit width was
0.4''; in the red arm, slit width was 0.3$\arcsec$. The spectral resolution
was about 110\,000 in the red and about 80\,000 in the blue. UVES data
were reduced using the automatic MIDAS UVES pipeline described by
Ballester et al.\ (2000).  Science frames were bias-subtracted and
divided by the extracted flat-field, except for the 860 setting, where
the 2D (pixel-to-pixel) flat-fielding was used, in order to better
correct for the fringing. Because of the high flux of the spectra we
use the UVES pipeline \textit{average extraction} method. 

The data can be accessed at {\tt http://archive.eso.org}.

\subsection{Previously published data}

In addition, all previously-published longitudinal magnetic field measurements of $\alpha$~And (Borra \& Landstreet (1980), Glagolevskij et al. (1985) and Chountonov (2001) have been recovered and considered in our analysis. These data are summarised in Table 2.

\setcounter{table}{1}

\begin{table}
\begin{center}
\begin{tabular}{ccr}
\hline
HJD & Phase  & $B_\ell\pm \sigma_B$ (G)\\
\noalign{\smallskip}
\hline
\noalign{\smallskip}
$^1$2443383.775 &  0.17923 &  $    35 \pm     55 $ \\
$^1$2443384.779 &  0.60066 &  $   118 \pm     44 $    \\
$^1$2443385.829 &  0.04140 &  $   -62 \pm     48 $    \\
$^1$2443387.812 &  0.87377 &  $     5 \pm     44 $    \\
$^1$2443388.813 &  0.29394 &  $   -38 \pm     55 $    \\
\noalign{\smallskip}
\hline
\noalign{\smallskip}
$^2$2445244.427 &  0.19134 &    $    31 \pm     76 $ \\
$^2$2445245.542 &  0.65936 &    $   -52 \pm    118 $ \\
$^2$2445303.427 &  0.95670 &    $   -62 \pm     41 $ \\
$^2$2445627.233 &  0.87486 &    $   -78 \pm     54 $ \\
$^2$2445629.242 &  0.71815 &    $   +64 \pm    112 $ \\
$^2$2445690.291 &  0.34357 &    $  -138 \pm     86 $ \\
\noalign{\smallskip}
\hline
\noalign{\smallskip}
$^3$2451798.542 & 0.29319 &    $  +20 \pm     50 $ \\
$^3$2451904.250 & 0.66432 &    $  -40 \pm     60 $ \\
   \hline\hline\noalign{\smallskip}
\end{tabular}
\caption[]{Archival longitudinal magnetic field measurements. $^1$Borra \& Landstreet (1980); $^2$Glagolevski et al. (1985); $^3$Chountonov (2001). Phases are calculated according to the ephemeris described in the text.}
\label{tab:journal}
\end{center}
\end{table}

\section{Ripple suppression in MuSiCoS Stokes $V$ spectra}

Upon reduction of the MuSiCoS spectra of $\alpha$ And, ripples were
apparent in both Stokes $I$ and Stokes $V$ with peak-to-peak
amplitudes of order 0.2\% and wavelengths of about 4~\AA\ in Stokes
$V$, and amplitudes of order 1\% and wavelengths of about 1~\AA\ in
Stokes $I$. Such ripples are present in most polarimeters (e.g. analogous ripples are detected in the ESPaDOnS Stokes $I$ spectra, although none are seen in the Stokes $V$ spectra), and
generally result from internal reflections producing secondary beams
coherent with the incident beam, but with important (wavelength
dependent) phase differences (Semel 2003). Such ripples are almost
certainly present in all MuSiCoS Stokes $V$ spectra, but because of the particular length scale and intensity of the
fringes, they can be effectively ignored in most MuSiCoS
applications. However, due to the (relatively) sparse line spectrum,
broad lines, and high polarimetric precision required for
$\alpha$~And, they must be removed (or at least suppressed) in this
case.

The Stokes $V$ ripple power spectrum for one of the spectral orders is shown in
Fig. 1. There are a number of periodicities, but most of the power is
at frequencies  $0.015-0.022\ {\rm pixel}^{-1}$. Both the amplitude
and the period of the ripples are functions of (spectrum) wavelength.
The period, or more accuractely the characteristic scale of fringes,
increases with wavelength from 3~\AA\ in the blue to 7~\AA\ in the
red (corresponding to an approximately constant wavelength in velocity units of 200~\kms).  Ripples in the blue part of the spectrum are characterised by
larger amplitudes.

   \begin{figure}[t]
   \centering
   \includegraphics[width=8.0cm]{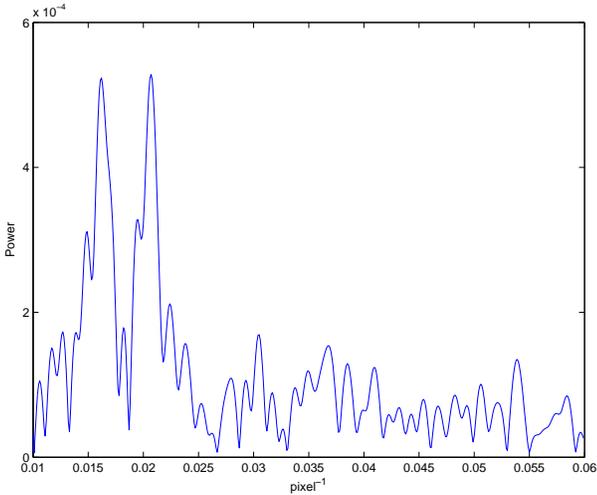}
         \label{}
\caption{Stokes $V$ ripple wavelet power spectrum, obtained using the spectrum of $\alpha$~Lyr. Note the evident power at scales 0.015 - 0.022 pixel$^{\rm -1}$ due to polarimetric ripples.}   
   \end{figure} 

   \begin{figure}[]
   \centering
   \includegraphics[width=9.0cm]{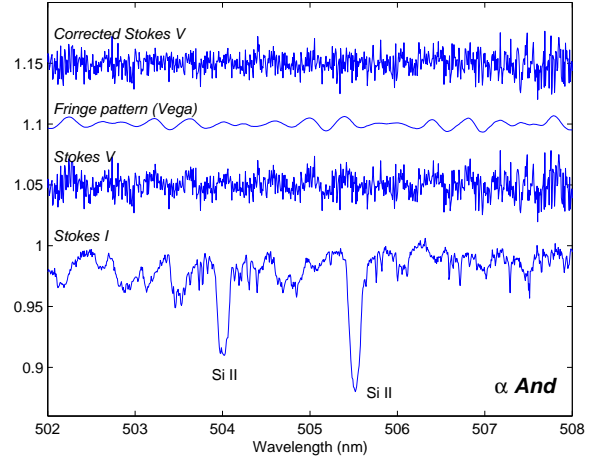}\vspace{0.25cm}
   \includegraphics[width=9.0cm]{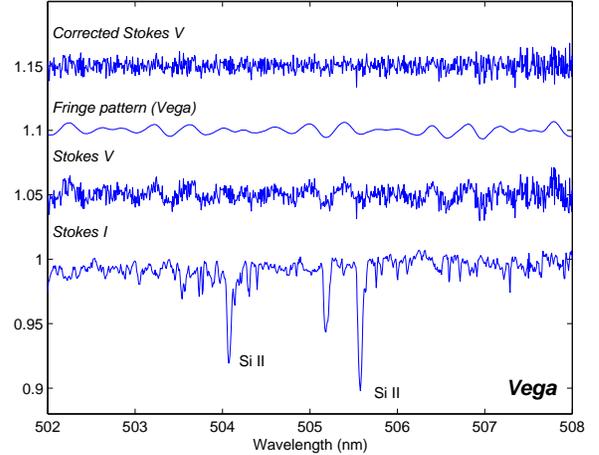}
         \label{}
\caption{Ripple removal from spectra of $\alpha$~And using Vega as a template.\ {\em Lower frame --}\ Vega\ {\em Upper frame -}\ $\alpha$ And. In both frames, raw and corrected Stokes $V$, as well as the fringe pattern, have been scaled and shifted for display purposes.}   
   \end{figure} 

Ripple removal was performed using a wavelet transform procedure {(e.g. Press at al. 1992)}.  This method allows one to isolate
and filter an aperiodic signal with a complicated interaction of
the amplitude, frequency and phase variation.  The signal is compared
not with an infinite sinusoid (as is done in the Fourier transform)
but with a finite pattern.  {\it Unlike} sines and cosines, wavelet
functions are localised in space.  Simultaneously, {\it like} sines
and cosines, wavelet functions are localised in frequency or (more
precisely) characteristic scale.  Hence, the wavelet transform
provides the possibility to isolate the contribution of a certain
scale, no matter what the complexity of the signal on that scale.

Polarisation features, if they exist in the Stokes $V$ spectrum of
$\alpha$\,And, have full widths similar to the period of the ripples
(as the projected rotational velocity $v\sin i=52$~\kms). Since
potential polarisation features and ripples have similar scales it is
hard to separate them.  However, similar ripple patterns are observed
in the Stokes $V$ spectrum of another, more slowly-rotating A star,
$\alpha$\,Lyrae (see Fig. 1), observed by Henrichs et al. during the 2003
MuSiCoS observing run.  We therefore decided to use this star as a
``fringing template''.  $\alpha$\,Lyr is a non-magnetic star, at least
at the level of about 10 G (e.g. Borra, Fletcher \& Poeckert 1981),
and we do not expect any polarisation features in its spectra.  Even
if they do exist, their frequency is about twice higher (the spectral
lines are narrower) than that of the fringes, and these features as
well as other high frequency components such as noise are filtered out
from the ripple pattern.  Ultimately, we used the ripple pattern
extracted from the $\alpha$\,Lyr reference spectra to correct our
spectra of $\alpha$\,And. As can be seen in Fig. 2, ripples are
significantly suppressed (although not completely removed) from the
Stokes $V$ spectra.  As a result, the S/N ratio in the corresponding
LSD profiles (see Sect. 4) increase by 10--15\%, and obvious evidence of ripples is
effectively removed.

\section{Extraction of LSD profiles}

Least-Squares Deconvolution (LSD; Donati et al. 1997, Wade et al. 2000a) was employed to extract high-precision mean Stokes $I$ and
$V$ Zeeman signatures from the MuSICoS and ESPaDOnS spectra. For the extraction we employed a
specially-designed HgMn line mask (see, e.g., Shorlin et al. 2002)
calculated for an effective temperature of 14000~K (Ryabchikova et
al. 1999). A total of 413 lines were used in the LSD of the MuSiCoS spectra, and { 582} lines in the LSD of the ESPaDOnS spectra { (in the latter, excluding lines significantly blended by telluric lines).}

LSD profiles obtained within $\pm0.05$ days of each other were averaged pixel-by-pixel, finally resulting in 18 sets of Stokes $I$ and $V$ LSD profiles (summarised in Table 3) that were employed for all of the following analysis. 

A careful examination of the MuSiCoS LSD profiles shows weak, residual ripples in some of the highest-S/N Stokes $V$ profiles, identifiable by a marginal but consistently higher circularly polarised flux in the red wing of the mean line. A full wavelength is apparent, and corresponds to about 200~\kms, in good agreement with the ripple wavelengths observed in the reduced spectra. The ripples, with full amplitudes of $1.6\times 10^{-4}$,  extend well outside the mean profile. The MuSiCoS and ESPaDOnS LSD Stokes $V$ profiles are illustrated in Fig. 3.

   \begin{figure}[t]
   \centering
   \includegraphics[width=7cm,angle=-90]{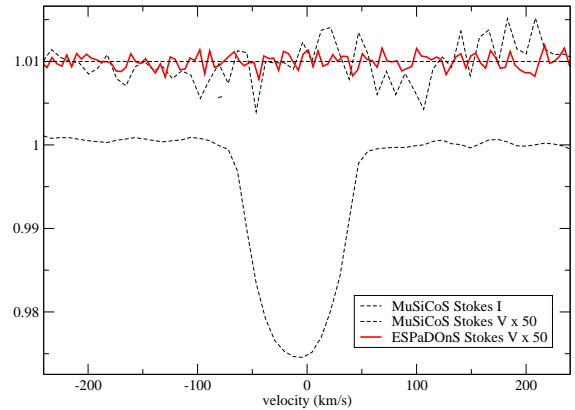}
         \label{}
\caption{Example MuSiCoS (9 Jul 03) and ESPaDOnS (21 Feb 05) LSD profiles. Note the residual ripples in the MuSiCoS Stokes $V$ profile, which extend well outside the range of the Stokes $I$ profile.}   
   \end{figure}

\begin{table}
\begin{center}
\begin{tabular}{rcc|rr|rr}
\hline
\# & Phase  & $N_{V}$ (\%)& $B_\ell\pm \sigma_B$& $|z_V|$ &$B_\ell\pm \sigma_B$& $|z_N|$\\
      &        &           & \multicolumn{2}{c|}{$V$} & \multicolumn{2}{c}{$N$}  \\
\noalign{\smallskip}
\hline
\noalign{\smallskip}
1 & 0.238 &   0.012  &  $    90 \pm     76 $&       1.2  &   $  -64 \pm     92 $ &     0.7\\
2 & 0.226 &   0.008  &  $    17 \pm     54 $&       0.3  &   $  -14 \pm     53 $ &     0.3\\
3 & 0.065 &   0.007  &  $   -85 \pm     44 $&       1.9  &   $   59 \pm     44 $ &     1.3\\
4 & 0.419 &   0.008  &  $    21 \pm     53 $&       0.4  &   $  -38 \pm     53 $ &     0.7\\
5 & 0.685 &   0.010  &  $   -33 \pm     64 $&       0.5  &   $  -13 \pm     63 $ &     0.2\\
6 & 0.522 &   0.004  &  $   -56 \pm     29 $&       2.0  &   $  -35 \pm     28 $ &     1.3\\
7 & 0.946 &   0.005  &  $   -70 \pm     34 $&       2.1  &   $   28 \pm     34 $ &     0.8\\
8 & 0.362 &   0.009  &  $   -72 \pm     58 $&       1.3  &   $   21 \pm     57 $ &     0.4\\
9 & 0.760 &   0.008  &  $   -34 \pm     52 $&       0.7  &   $  -87 \pm     51 $ &     1.7\\
10& 0.863  &   0.005  & $    -89 \pm     31$ &       2.9  &  $   -11 \pm     30$  &     0.4\\
11& 0.283  &   0.006  & $    -62 \pm     39$ &       1.6  &  $    35 \pm     38$  &     0.9\\
12& 0.699  &   0.005  & $    -59 \pm     33$ &       1.8  &  $    36 \pm     32$  &     1.1\\
13& 0.658  &   0.005  & $    -62 \pm     34$ &       1.8  &  $     3 \pm     33$  &     0.1\\
\noalign{\smallskip}
\hline
\noalign{\smallskip}
14& 0.459  &  0.0013  & $   +2 \pm     8$  &       0.3  &  $    +5 \pm     8$  &     0.6\\
15& 0.315  &  0.0033  & $   -9 \pm    19$ &        0.5  &  $    -30 \pm     19$  &   1.6\\
16& 0.404  &  0.0010  & $    -2 \pm    6 $ &       0.3  &  $      1 \pm      6$  &   0.2  \\
17& 0.808  &  0.0010  & $     2 \pm    6 $ &       0.3  &  $     -3 \pm      6$  &   0.5  \\
18& 0.240  &  0.0010  & $    -4 \pm    6$ &        0.7  &  $      7 \pm      6$  &   1.2  \\
   \hline\hline\noalign{\smallskip}
\end{tabular}
\caption[]{Final averaged LSD Stokes $I$ and $V$ profile sets and their corresponding inferred longitudinal magnetic fields. Profiles 1-13 correspond to MuSiCoS spectra; profiles 14-18 corresponds to ESPaDOnS spectra. Phases correspond to the time of mid-observations, and are calculated according to the ephemeris described in the text. $N_V$ is the noise level in the Stokes $V$ profile. Longitudinal fields are measured from both the Stokes $V$ and diagnostic null spectra $N$, and are reported (along with their associated detection significance $z$) in the final 4 columns.}
\label{tab:journal}
\end{center}
\end{table}

\section{Constraints on the photospheric magnetic field}

\subsection{Longitudinal magnetic field measurements}

\subsubsection{MuSiCoS and ESPaDOnS LSD profiles}

   \begin{figure}[t]
   \centering
\includegraphics[width=7cm,angle=-90]{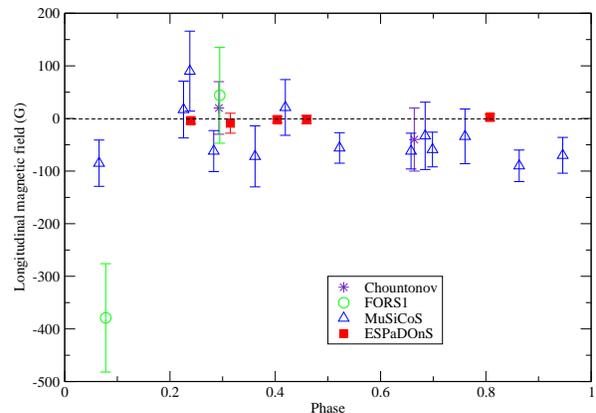}
          \label{}
\caption{Longitudinal magnetic field measurements (epoch 3) as a function of rotational phase. }
   \end{figure}

Longitudinal magnetic fields were inferred from each of the Stokes $I$ and $V$ profile sets, using the first-order moment method. According to this method, the longitudinal field $B_\ell$ (in gauss) is calculated from the Stokes $I$ and $V$ profiles in velocity units as:
\begin{equation}
B_{\ell} = -2.14 \times 10^{12} \frac{\int_{}^{}vV(v)dv}{\lambda zc
\int_{}^{}[I_c - I(v)]dv},
\end{equation}

\noindent (Donati et al. 1997; Wade et al. 2000b) where $\lambda$, in \AA, is the mean wavelength of the LSD profile, $c$ is the velocity of light (in the same units as $v$), and $z$ is the mean value of the Land\'e factors of all
lines used to construct the LSD profile. The accuracy of this
technique for determining high-precision longitudinal field
measurements has been clearly demonstrated by Wade et al. (2000b),
Donati et al. (2001) and Shorlin et al. (2002).

 The resultant MuSiCoS longitudinal magnetic field measurements have associated 1$\sigma$ uncertainties from 29-76 G, and a median uncertainty of 44 G. Although we detect no significant longitudinal magnetic field variation, the majority of the measurements are negative, typically falling 1-2$\sigma$ below zero, with a mean of $-38\pm 12$~G. No similar bias is observed in the longitudinal field measurements inferred using the LSD diagnostic $N$ profiles. 

 The new LSD $B_\ell$ measurements are listed in Table 3.

\subsubsection{FORS1 spectra}

Longitudinal magnetic fields were inferred from the two FORS1 spectra of $\alpha$~And using full-spectrum regression according to the expression:

\begin{equation}
\Bigl(\frac{V}{I}\Bigr) = -g_\mathrm{eff} \ \Delta \lambda_z \ \lambda_0^{2} \
              \frac{1}{I} \
              \Bigl(\frac{{\partial}I}{\partial\lambda}\Bigl) \
              B_\ell
\end{equation}

\noindent (Bagnulo et al. 2002, 2005). Here, $g_\mathrm{eff}$ is the Land\'e factor (equal to 1.0 for Balmer lines, and assumed equal to 1.25 for metal lines), $\Delta\lambda_z$ is the Lorentz unit, $\lambda_0$ is the line wavelength, { $I$ is observed flux at wavelength $\lambda$}, and $B_\ell$ is the longitudinal magnetic field. As described by Bagnulo et al. (2002), the longitudinal field is obtained by a Least-Squares straight-line fit to the regression of $V/I$ versus $-\Delta \lambda_z \ \lambda^{2}$ ({1}/{$I$}) ($\partial I$/$\partial \lambda$). The inferred values of $B_\ell$, obtained using the entire spectrum (metal + Balmer lines), are reported in Table 4. Notably, one of the FORS1 measurements corresponds to a 3.7$\sigma$ detection ($B_\ell=-379\pm 103$~G).

\begin{table}
\begin{center}
\begin{tabular}{ccr}
\hline
HJD & Phase  & $B_\ell\pm \sigma_B$ (G)\\
\noalign{\smallskip}
\hline
\noalign{\smallskip}
2452910.592 &  0.07823 &  $  -379 \pm    103 $ \\
2452963.520 &  0.29486 &  $   +44 \pm     91 $    \\
   \hline\hline\noalign{\smallskip}
\end{tabular}
\caption[]{Longitudinal magnetic field of $\alpha$~And inferred from FORS1 measurements.}
\label{tab:journal}
\end{center}
\end{table}

\subsection{Modeling the phased $B_\ell$ measurements}

Of the 33 longitudinal field measurements, the best uncertainty (1$\sigma$) corresponds to three of the ESPaDOnS measurement ($6$~G), whereas the worst uncertainty corresponds to one of the measurements of Glagolevski et al. (1985; $-52\pm 118$~G), a range of almost a factor of 20 in precision! The median uncertainty of the measurements is 45 G. 

The longitudinal field measurements span a total time of about $10^4$ days. Unfortunately, the $\pm 0.00011$~d uncertainty associated with the rotational period of Adelman et al. (2002) produces a phase uncertainty of the oldest measurements relative to the newest measurements equal to $\pm 0.19$ cycles. 

All 33 $B_\ell$ measurements (previously published, LSD and FORS1) have been assigned phases according to the rotational ephemeris of Adelman et al. (2002):

\begin{equation}
{\rm JD} = 2449279.6898 + (2.38236\pm 0.00011)\cdot E.
\end{equation}

\noindent However, as indicated above, the relative phases of $\alpha$~And at the various epochs of observation are uncertain by up to $0.19\times 2\approx 0.4$ cycles. The newer and older longitudinal field data therefore cannot be phased together, and must be analysed separately.


As a first step to constrain the presence of magnetic fields in the photosphere of $\alpha$~And, we have modelled the phased longitudinal field measurements at three epochs: epoch 1 (HJD 2443383-2443388, measurements of Borra \& Landstreet 1980, internal phase uncertainty of $1\times 10^{-4}$ cycles), epoch 2 (HJD 2445244-2445690, measurements of Glagolevski et al. 1985, internal phase uncertainty of $\pm 0.008$ cycles), and epoch 3 (HJD 2451798-2453423, measurements of Chountonov (2001), as well as the MuSiCoS, FORS1 and ESPaDOnS measurements reported in this paper, internal phase uncertainty of $\pm 0.03$ cycles). The phased epoch 3 measurements are illustrated in Fig. 4. 

{ We begin by performing least-squares fits to the data. We find that the variable-field hypothesis (first-order sine fit, fitting zero-point, amplitude and phase), constant-field
hypothesis (i.e. a straight line through the weighted mean), null-field hypothesis (i.e. a straight line through $B_\ell = 0$) are indistinguishable (at the 99\% confidence level\footnote{Confidence intervals were obtained using $\chi^2$ probabilities from Press et al. (1992). According to these tables, the 99\% confidence interval, considering three model parameters, corresponds to an increase in total $\chi^2$ of 11.3.}) for each of the 3 epochs. We therefore conclude that no magnetic field is detected in the longitudinal field measurements of $\alpha$~And. }



To quantify the magnetic field upper limit implied by the measurements, we used a modified version of the programme {\sc Fldcurv} to calculate model longitudinal field variations corresponding
to a large number of dipolar magnetic field configurations, varying
the dipole (polar) intensity $B_{\rm p}$ as well as the obliquity angle
$\beta$ and the phase of meridian crossing of the positive magnetic pole $\phi_0$ (which corresponds to the phase of peak positive longitudinal field). {\sc Fldcurv} computes the surface distribution of magnetic field corresponding to a particular magnetic geometry,
 then weights and integrates the longitudinal component over the hemisphere of the star visible at a given phase. This procedure is repeated for a series of phases, producing a synthetic longitudinal field variation corresponding to the model geometry. For this procedure we have assumed a limb-darkening coefficient of 0.4.

For each model { (corresponding to a grid computed with 10~G, $1\degr$ and $0.05$ sampling of $B_{\rm d}, \beta$ and $\phi_0$, respectively)} the reduced $\chi^2$ of
the model fit to the epoch 3 observations was calculated (we select the epoch 3 observations because they will provide the best constraint, due to the ultra-high precision ESPaDOnS data). This analysis indicates that for
a rotational axis inclination $i=74\pm 10\degr$ (Ryabchikova et al. 1999), acceptable magnetic models (again, at the 99\% confidence level for epoch 3)
are characterised by $|B_{\rm p}|<225$~G. Both $\beta$ and the phase of peak longitudinal field $\phi_0$ are not constrained at this confidence level (i.e. $\beta$ and $\phi_0$ are formally free to adopt any value between 0 and $180\degr$, and 0.0 and 1.0, respectively). Results for epochs 1 and 2 provide qualitatively similar results, with larger upper limits due to the lower precision of their respective measurements.




\subsection{Modeling the phased Stokes $V$ profiles}

To more robustly search for photospheric magnetic
fields, we have also modelled the LSD mean Stokes $V$ profiles. To begin, we find that no Stokes $V$ signatures are detected within the boundaries of any of the 18 final individual LSD line profiles (according to the detection criterion described by Donati et al. 1997). Several of the highest-S/N MuSiCoS profiles show residual ripples which, as mentioned earlier, extend well beyond the profile boundaries. 

We have calculated synthetic Stokes $I$ and $V$ LSD
profiles, using the polarised line synthesis code Zeeman2 (Landstreet
1988, Wade et al. 2001), assuming a triplet Zeeman pattern and the mean Land\'e factor and wavelength appropriate to the $\alpha$ And line mask (the accuracy of such a modelling procedure has been demonstrated by e.g. Donati et al. 2001, 2002).  Using this line model, the phase-independent LSD Stokes $I$ profile was fit, varying the line depth, projected rotational velocity\footnote{{ This allows for an { independent} determination of $v\sin i=52\pm 2$~\kms.}} and radial velocity. We then performed calculations of Stokes $V$ signatures for a large grid of dipolar magnetic field models,
and compared the Stokes $V$ profile rotational variation predicted by each model with the observed LSD Stokes $V$ phase variation. { The model grid was computed with sampling of 25~G, $5\degr$ and $0.05$ cycle sampling of $B_{\rm d}, \beta$ and $\phi_0$, respectively)}

We have performed independent analyses of the MuSiCoS and ESPaDOnS profiles.{ The results are derived from examination of concordance maps, examples of which are shown in Fig. 5. The concordance maps show the summed reduced $\chi^2$ for each model/data comparison as a function of the model magnetic
field polar strength $B_{\rm p}$ and obliquity $\beta$. White regions correspond to models which are {\em inconsistent} with the observations at more than 99\% confidence, whereas dark regions correspond to models that agree with the observations. }

   \begin{figure*}[th]
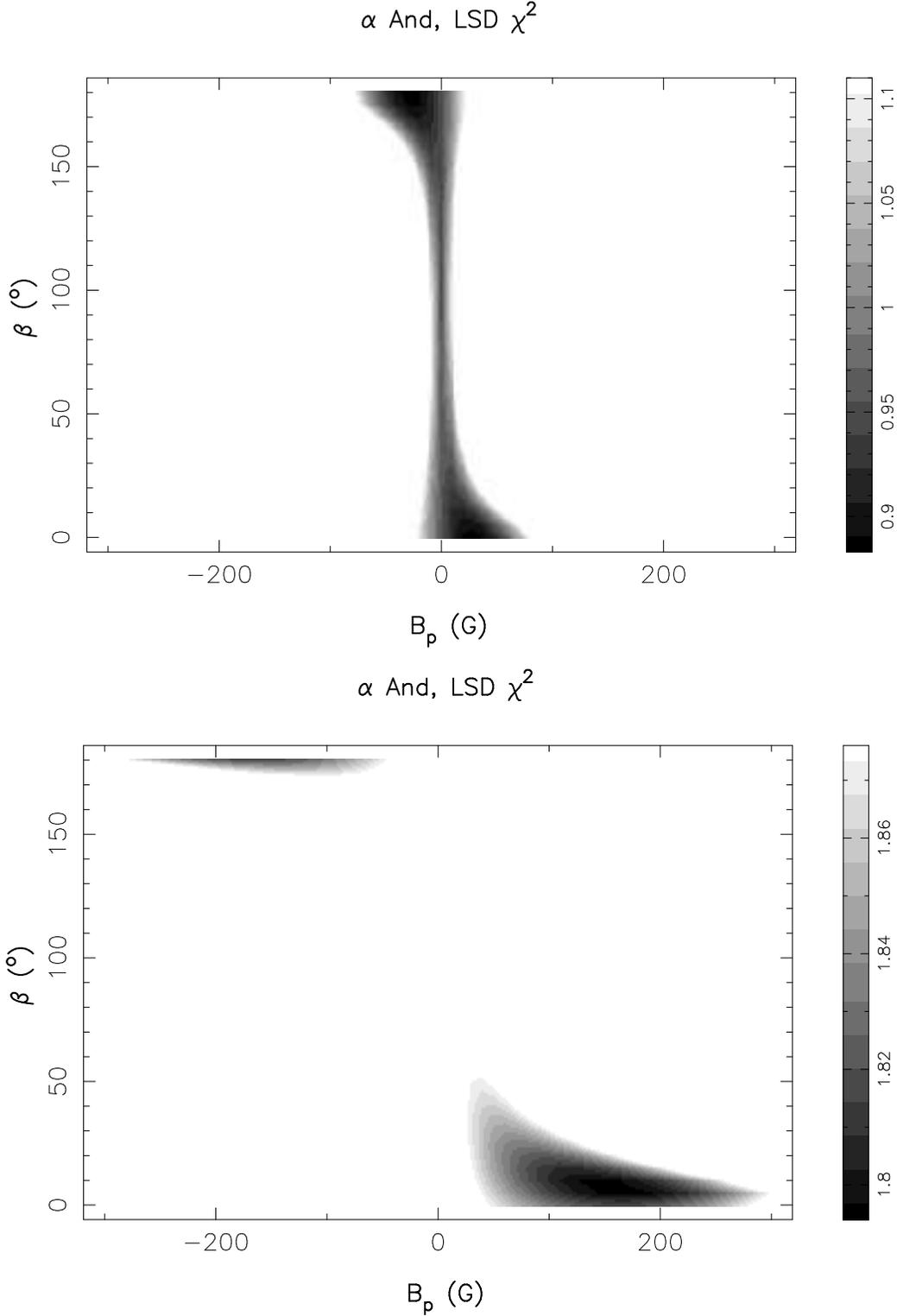

   \centering
   \includegraphics[width=10.0cm,angle=-90]{espadons_chi2.ps}\vspace{0.25cm}
   \includegraphics[width=10.0cm,angle=-90]{musicos_chi2.ps}\vspace{0.25cm}
         \label{}
\caption{Reduced $\chi^2$ of the Stokes $V$ model/data comparison for dipole fields, as a function of the polar field strength $B_{\rm p}$ and the obliquity between the rotational axis and the magnetic field symmetry axis $\beta$. A rotational axis inclination of $i=74^{\rm o}$ has been assumed for all models. Dark regions correspond to models consistent with the observations, while white regions correspond to models incompatible with the observations at more than 99\% confidence. {\em Upper frame -}\ ESPaDOnS data ($\phi_0=0.00$) \ {\em Lower frame\ -}\ MuSiCoS data ($\phi_0=0.00$).}   
   \end{figure*} 

For the ESPaDOnS data { (upper panel of Fig. 5)}, allowing the three parameters to vary freely, we obtain (at 99\% confidence) that $|B_{\rm p}|<90$~G. Moreover, for all significantly oblique dipolar magnetic field geometries ($30\degr<\beta<150\degr$), we obtain a substantially stronger constraint: $|B_{\rm p}|<40$~G. { For the ESPaDOnS profiles, both $\beta$ and $\phi_0$ are unconstrained, and can adopt any value between $0-180\degr$ and $0.0-1.0$, respectively. }


For the MuSiCoS profiles { (lower panel of Fig. 5)}, the results are very different. We obtain 99\% confidence intervals of $|B_{\rm p}|=25-300$~G, $\beta=0-70\degr$ $\phi_0=0.0-1.0$, and best-fit values of $|B_{\rm p}|=150$~G and $\beta=10\degr$. Using {\sc Fldcurv} to model the phased MuSiCoS longitudinal field 


\noindent measurements alone, we also obtain significantly non-zero values for $|B_{\rm p}|$ { ($250-1100$~G), and $\beta=0-20\degr$.}


Finally, we have also calculated a series of pure quadrupolar and octupolar magnetic field models, in order to test the ESPaDOnS observations' sensitivity to more complex field configurations. For both pure quadrupolar and octupolar fields, we find that the polar field $|B_{\rm p}|<110$~G (and $<50$~G for the majority of geometries).

\subsection{Interpretation of the magnetic field results}

The analysis carried out in Sect. 5.3 indicates that the combined magnetic data provide no evidence for the presence of magnetic fields in the atmosphere of $\alpha$~And. Moreover, the ultra-high precision ESPaDOnS measurements constrain any low-order multipolar magnetic field to be weaker than about 100~G.

Nevertheless, both the MuSiCoS data and one of the FORS1 measurements are not fully consistent with the general conclusion of the absence of a magnetic field. As discussed in Sect. 5.1.1, the MuSiCoS longitudinal field measurements show marginally but systematically negative values, and modelling of these data and the associated LSD Stokes $V$ profiles suggests the presence of a weak magnetic field (150~G) nearly aligned with the rotational axis of the star ($\beta=10\degr$). Such a magnetic field geometry is characterised by Stokes $V$ signatures of constant sign, and nearly constant shape and intensity. However, we have also identified residual ripples in the MuSiCoS LSD profiles. Because the residual ripples are also of constant sign and approximately constant shape and intensity, it seems rather likely that the model is fitting the residuals rather 

\clearpage

\noindent than any real circular polarisation. This is confirmed by a detailed examination of the highest-S/N data and the best-fit model, which reveals a close correspondence between the ripple residuals and the predicted Stokes $V$ signatures. Thus we conclude with confidence that the apparent detection of a field from the MuSiCoS measurements results directly from incomplete suppression of interference ripples.

Although none of the individual MuSiCoS longitudinal field measurements corresponded to a significant detection, one of the FORS1 measurements corresponds to a field detected at 3.7$\sigma$. In order to evaluate the reliability of this marginal detection, we have performed a detailed re-analysis of the spectrum, including an examination of the contribution of individual frames and spectral regions (e.g. Bagnulo et al. 2005). First, we find no significant magnetic field in the metallic line spectrum (i.e. excluding the Balmer lines from the analysis: $B_\ell=-485\pm 210$~G), although with a relatively large uncertainty. Secondly, we find that all frames contribute more or less equally to the detection, and there is no indication that a particular subset of the frames is responsible for the detection. Finally, an examination of the longitudinal fields inferred from individual Balmer lines shows a diversity of results. None of the individual Balmer lines shows a significant field detection, although the combined analysis of Balmer lines with indices larger than H9\footnote{Where index 9 refers to the quantum number of the upper level in the transition.} does produce a marginally significant ($3.3\sigma$) field ({ notably} at the level of nearly 1 kG). Most of the individual Balmer lines do show systematically negative values of the longitudinal field.


Based on this analysis alone, there is no reason to reject the apparent FORS1 detection. However, it is curious that other independent longitudinal field measurements, obtained at phases close to that of the FORS1 detection, are consistent with zero field (e.g. $-85\pm 44$~G at phase 0.065; see Fig. 4). One explanation for the apparent detection may be related to the very short exposure duration required to observe $\alpha$~And with FORS1, which is comparable to the time required to open the shutter. So, although we are unable to confidently explain the single FORS1 detection, this measurement should be interpreted in the context of the full data set (in particular as, even in the absence of a magnetic field, gaussian statistics indicates the probability of a single detection between 3 and 4$\sigma$ in a sample of 33 measurements to be 9\%).

{ The potential for spurious detection of fields, in particular due to ripples in MuSiCoS Stokes $V$ spectra, deserves further comment. In the case of $\alpha$~And, ripples in the Stokes $V$ spectrum yielded a small but measurable contribution to the longitudinal field and the LSD Stokes $V$ signatures due to the very high S/N of the spectra, the relatively close correspondence between the ripple wavelength and the stellar line width, and the relatively small number of lines employed in the LSD. It is the {\em combination}s of these conditions that produces a situation where ripples could potentially affect the magnetic field diagnosis. A similar apparent contribution of ripples to the LSD Stokes $V$ profile is reported by Shorlin et al. (2002) for the hot Am star HD 89021 (S/N=910, $v\sin i=50$~\kms, $T_{\rm eff}\simeq 9000$~K), for which essentially the same conditions are satisfied. Fortunately, the presence of significant ripple contributions can be ruled out in most cases (and in particular in essentially all observations of magnetic Ap stars and cool stars) either by looking for { small-amplitude} structures in the LSD profile extending outside of the line profile which do not vary significantly with phase, or using observations of magnetic and non-magnetic standards. { The absence of any known examples of such phenomena, apart from HD~89021 and $\alpha$~And, suggests that detectable ripple contributions are extremely rare (given that several hundred different stars have been observed using this instrument).}}

Retaining therefore our conclusion that $\alpha$~And is a non-magnetic star, what are the implications of our magnetic field upper limits? In the atmosphere of $\alpha$ And, at $\log \tau_{\rm 5000}=0.0$, the magneto-thermodynamic equipartition field strength (for
which the magnetic and thermodynamic energy densities are
approximately equal) is $B_{\rm eq} = \sqrt{12\pi nkT}\simeq 265$~G.
This is significantly (2.7-6.5 times) larger than the (3$\sigma$) upper limits derived from modelling of the ESPaDOnS Stokes profiles for any low-order multipolar field that might be present. 

   \begin{figure*}[t]
   \centering
   \includegraphics[width=7cm,angle=-90]{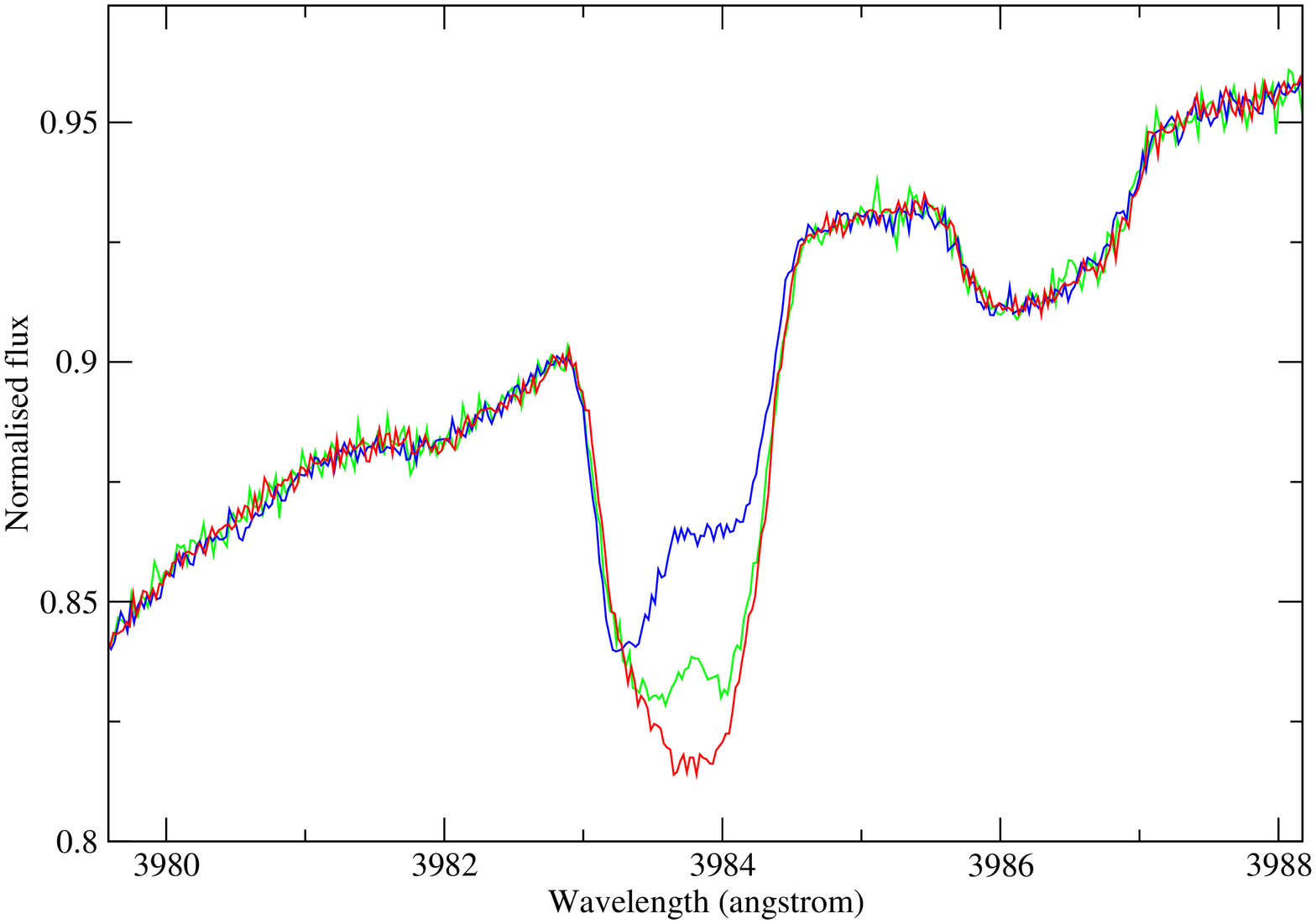}\includegraphics[width=7cm,angle=-90]{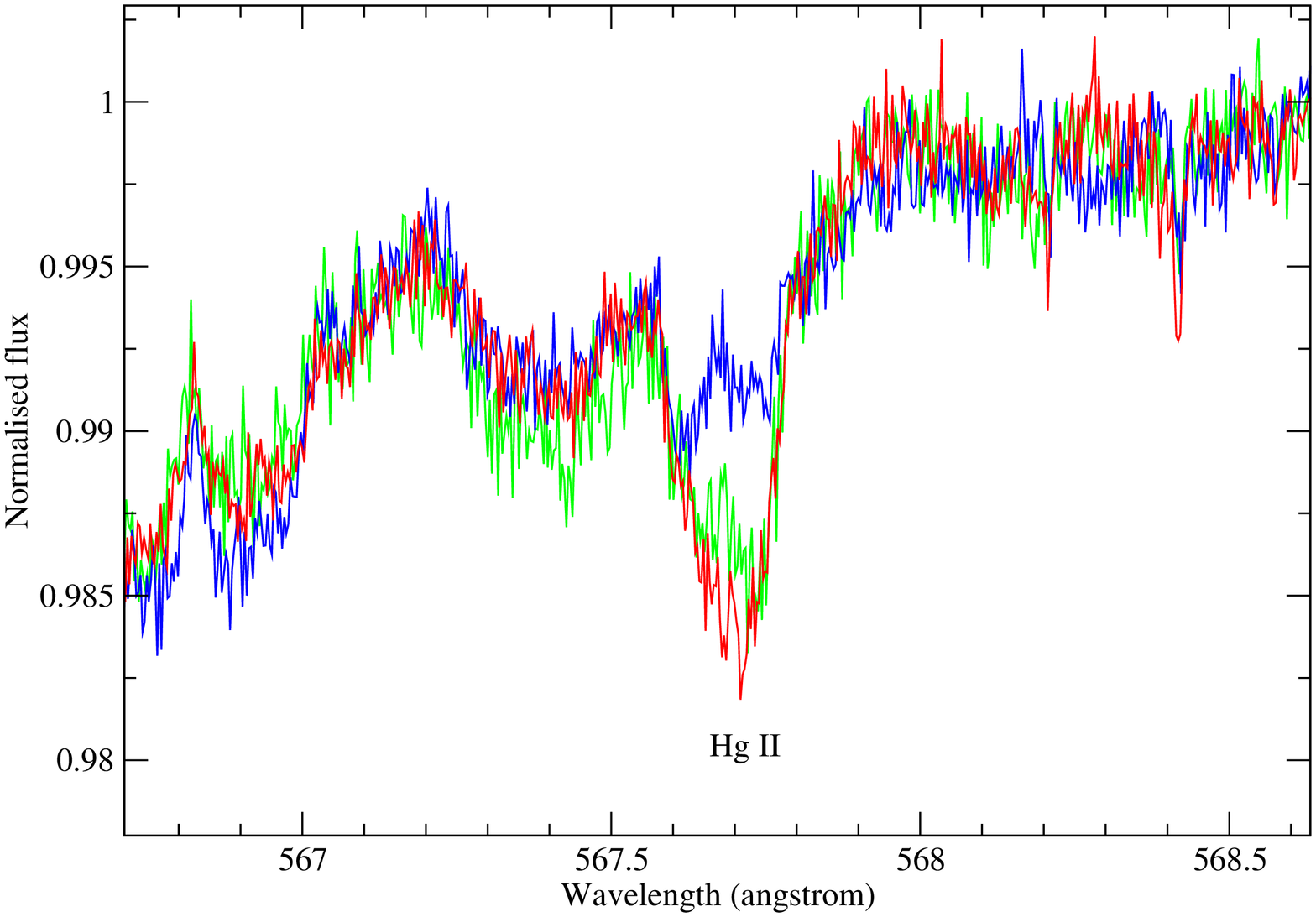}
         \label{}
\caption{Variability of Hg~{\sc ii} 3984 and Hg~{\sc ii} 5677 in ESPaDOnS spectra of $\alpha$~And.}
   \end{figure*}

Although we have performed no quantitative modeling of
more complex magnetic structures in this study (fields such as those exhibited
by active late-type stars like HR 1099, e.g. Petit et al. 2004a, for example),
the fact that relatively strong, complex fields are easily detected in LSD profiles of
similar quality in a sample of late-type stars of similar $v\sin i$ using the MuSiCoS and ESPaDOnS
spectropolarimeters strongly suggests that similar fields are not present in the
photosphere of $\alpha$ And. For example, Petit et al. (2004b) find Stokes $V$ amplitudes of
0.05-0.1\% due to a complex surface magnetic field in spectra of the rapidly-rotating ($v\sin i=70$~\kms)
FK Com star HD 199178. Such signatures would be easily detectable in our LSD profiles of $\alpha$~And (either the MuSiCoS or ESPaDOnS profiles), and we can therefore rule out the presence of such complex fields as well. 

Finally, although the ESPaDOnS measurements are not sufficiently sensitive to detect much weaker and/or more finely structured fields (such as those observed in the solar atmosphere), the total absence of any of the proxy phenomena well-known to be associated with such fields (spectroscopic emission lines, non-thermal radio emission, X-ray emission, flares, etc.) allows us to reasonably rule out their presence as well.

\section{A search for additional line profile variability}

The high quality of the MuSiCoS, ESPaDOnS and UVES spectra of $\alpha$~And
makes them well-suited for a search for line profile variations
similar to those reported by Adelman et al. (2002). The spectra were first searched at all locations of potential lines of Hg {\sc i, ii}\ and\ {\sc iii}, as predicted using the {\sc vald} database (Piskunov et al. 1995). Secondly, the spectra were visually examined throughout the entire observed domain, searching for any evidence of line profile variations or unusual line shapes that might be attributed to nonuniform abundance distributions. We point out that spectral variability due to the SB2 nature of this system makes confident identification of such phenomena very challenging.

The Hg~{\sc ii} $\lambda3984$ line in the UVES and ESPaDOnS spectra shows strong asymmetry and variability, and should be contrasted to nearby lines of Fe~{\sc ii} (Fig. 6, left panel). Thus we confirm the detection of variability in Hg~{\sc ii} $\lambda3984$ as reported by Ryabchikova et al. (1999) and Adelman et al. (2002). We also observe variability of Hg~{\sc ii} $\lambda 6149$, $\lambda 5425$ and $\lambda 5677$. Although this variability is not confidently seen in the UVES or MuSiCoS spectra (due to lower S/N), the very high S/N ESPaDOnS spectra show the variability quite clearly (Fig. 6). In particular, the variability is evident in the spectra obtained on the successive nights from Aug 23-25 2005. 

We also observe weak variability of other lines in the spectrum of $\alpha$~And. We have looked in some detail at these variations, which are present in nearly all regions of the spectra, and are quite subtle. A careful examination of the observed spectral variations on Aug 23-25 2005, in the light of synthetic spectra computed including the spectroscopic contribution of the secondary (kindly provided by T. Ryabchikova using the results of Ryabchikova et al. (1999), and taking into account the appropriate radial velocity at the time of each observation) suggests that variability of most lines can be approximately explained by the combination of spectroscopic variability due to orbital motion, small continuum normalisation differences, and weak, variable fringing. This conclusion is consistent with the observed diversity of variability characteristics of most elements (as opposed to the systematic variability observed in the Hg lines). 


\section{Discussion}

Until recently, $\alpha$ And represented the only convincing example of
a HgMn star showing intrinsic line profile variability. However, Kochukhov et al. (2005) have reported asymmetry and variability of the Hg~{\sc ii} $\lambda3984$ line in spectra of two additional HgMn stars, with temperatures similar to that of $\alpha$~And. This appears to establish a new class of spotted, spectrum variable stars, of which $\alpha$~And is the prototypical example.

From our analysis of the magnetic and polarisation measurements of $\alpha$~And, we find that any oblique low-order multipolar magnetic field present in the photosphere must be {weaker} than about 100~G (for all field geometries), and weaker than about 50~G (for geometries with significant obliquity). Any undetected field is therefore constrained to be significantly weaker than the equipartition threshold at $\log \tau_{\rm 5000}=0.0$. If we assume that a magnetic field of at least this
order is required to support horizontal abundance
nonuniformities { (e.g. by suppressing mixing due to meridional circulation)}, then {an oblique low-order multipolar magnetic field does not appear to be responsible
for the nonuniform distributions of Hg and Mn of $\alpha$ And,} at least in atmospheric
layers around $\log \tau_{\rm 5000}=0.0$. 
Observational support for this { assumption} is provided by Auri\`ere et al. (2004), who have concluded that all magnetic Ap/Bp stars host dipolar magnetic field components that are {stronger} than the photospheric
equipartition field (at $\log \tau_{\rm 5000}=0.0$). This supports the assumption that a field of at least this strength is required for magnetic maintenance of photospheric chemical abundance non-uniformities. 

Could it be that a dipolar magnetic field does in fact exist in the photosphere of $\alpha$~And, but which
is weaker than the local equipartition field in layers around $\log
\tau_{\rm 5000}=0.0$? Due to the exponential decrease of
the gas density with height in the outer layers of the star, such a
field could in fact be {stronger} than the equipartition field in
atmospheric layers well above (say) $\log \tau_{\rm 5000}=-1$. If the abundance non-uniformities were confined to
these high layers, an undetected magnetic field (i.e. with polar intensity weaker than our 3$\sigma$
upper limit) could in principle be responsible { for preventing horizontal mixing of the non-uniform
 abundance distributions.}

The concentration of Hg, Mn and other elements in the photospheres of
HgMn stars in thin layers at small optical depths has been suggested
based on both theoretical and observational grounds (e.g. Proffitt et
al. 1999,  Sigut et al. 2000). Most recently, the
discovery of emission lines in the optical spectra of some HgMn stars
(e.g. Sigut et al. 2000, Wahlgren \& Hubrig 2000) has provided
additional evidence of stratification of these elements. Sigut (2001),
interpreting the presence of Mn~{\sc ii} emission lines in the spectra
of HD 186122 and HD 179761 as the result of interlocked non-LTE
effects combined with chemical stratification, found that Mn must be
confined to column masses above $\log \Sigma=-1.5$, equivalent to
about $\log\tau_{\rm 5000}\ltsim -1.1$. Although we observe no such emission lines in the spectrum of $\alpha$~And, were Hg confined to very small optical depths in the atmosphere of this star (i.e. optical depths at which the equipartition field satisfies our field constraints - for $\alpha$~And, $B_{\rm eq}$ should be below 100~G above $\log \tau_{\rm 5000}=-1.2$, and below 50~G above $\log \tau_{\rm 5000}=-2.2$) it might be
possible for an undetected magnetic field to { sustain} the lateral
abundance non-uniformities.

In order to test this possibility, {we have performed some comparisons between the observed Hg line profiles and those predicted by a single-step stratified Hg distribution, in which Hg is 100 times more abundant above $\log\tau_{\rm 5000}=-1.2$ than below, and in which the lower zone abundance { (we employed $\epsilon_{\rm Hg}=-5.0$)} has been chosen to approximately fit the profile of Hg~{\sc ii}~$\lambda 3984$. Unfortunately, the combination of line asymmetry and variability, coupled with the weakness of many of the Hg lines (1-2\% deep) and local continuum normalisation errors, makes it impossible to realistically choose between the stratified and unstratified models. }

{ Ultimately, we are led to conclude that models seeking to explain the nonuniform Hg surface distribution of $\alpha$~And which invoke the presence of a magnetic field of $B\gtrsim 100$~G are almost certainly falsified by the results presented in this paper. It may still be possible to implicate a dynamically-important magnetic field if Hg is strongly stratified, and confined primarily to upper layers of the atmosphere (where the equipartition field is weaker than the upper limits presented here). Based on our experiments, modeling of the vertical Hg distribution will require very high quality data (of quality similar to or better than that discussed here), coupled with a model which simultaneously considers the lateral and vertical abundance distribution. However, in the absence of evidence of strong Hg stratification, we conclude that { a non-magnetic} mechanism is almost certainly responsible for this phenomenon { (see e.g. Kochukhov et al. 2005).}}


\begin{acknowledgements}
Sincere thanks to H. Henrichs for the use of unpublished spectra of $\alpha$ Lyr employed for fringe suppression in this study. GAW extends warm thanks to T. Ryabchikova for her model computations and extremely valuable advice and discussions.

GAW and JDL acknowledge Discovery Grant support from the Natural Sciences and Engineering Research Council of Canada. 

This paper has benefitted significantly from discussions with O. Kochukhov and T.A.A. Sigut. Thanks also to G. Chountonov for providing details of his polarimetric observations of $\alpha$~And. 

\end{acknowledgements}
\end{document}